# Dynamic instability of individual carbon nanotube growth revealed by *in situ* homodyne polarization microscopy

**Authors:** Vladimir Pimonov[1], Huy-Nam Tran[1], Léonard Monniello[1], Saïd Tahir[1], Thierry Michel[1], Renaud Podor[2], Michaël Odorico[2], Christophe Bichara,[3] Vincent Jourdain[1*]

**Affiliations:**

[1] Laboratoire Charles Coulomb (L2C), UMR 5221 CNRS-Université de Montpellier, Place Bataillon, Montpellier, FR-34095, France.

[2] ICSM, Univ Montpellier, CEA, CNRS, ENSCM, Bagnols sur Cèze, FR-30207 Cedex, France.

[3] Aix Marseille Univ, CNRS, Centre Interdisciplinaire de Nanoscience de Marseille, Marseille, FR-13288 Cedex 09, France

*Correspondence to: vincent.jourdain@umontpellier.

**Abstract**:

Understanding the kinetic selectivity of carbon nanotube growth at the scale of individual nanotubes is essential for the development of high chiral selectivity growth methods. Here we demonstrate that homodyne polarization microscopy can be used for high-throughput imaging of long individual carbon nanotubes under real growth conditions (at ambient pressure, on a substrate), and with sub-second time resolution. Our *in situ* observations on hundreds of individual nanotubes reveal that about half of them grow at a constant rate all along their lifetime while the other half exhibits stochastic changes in growth rates, and switches between growth, pause and shrinkage. Statistical analysis shows that the growth rate of a given nanotube essentially varies between two values, with similar average ratio (~1.7) regardless of whether the rate change is accompanied by a change in chirality. These switches indicate that the nanotube edge or the catalyst nanoparticle fluctuates between different configurations during growth.

**Main:**

With their high carrier mobility, low capacitance and high chemical stability, carbon nanotubes (CNTs) are theoretically ideal for building highly-scaled transistors delivering large drive currents at low energy consumption [1]. The standard method for growing CNTs is catalytic chemical vapor deposition (C-CVD) in which nanoparticles (*e.g.*, Fe, Ni) catalyze the decomposition of a gaseous carbon precursor and the assembly of carbon atoms in a one-dimensional tubular crystal [2]. A key issue is that CNT samples are commonly made of different chiralities having different metallic (M) or semiconducting (SC) type and bandgap energies [3]. Remarkable advances in the selective growth of CNTs with specific metallicity or chirality have been empirically made during the last years [4–8]. However, the origin of this selectivity is still a matter of controversy, thus hindering the rational development of more selective strategies. In particular, the understanding of the processes governing CNT growth kinetics and their dependence on chirality and extrinsic factors (*e.g.*, catalyst, gas species) is still very limited.

For CNT ensembles, experimental measurements generally display an exponential decay of the growth rate [9–12]. Puretzky *et al.* proposed a kinetic model explaining this by a progressive encapsulation of the catalyst by a carbonaceous layer [13]. Yakobson *et al.* proposed different models predicting a constant growth rate but with different chirality dependence: proportional to the chiral angle $\chi$ [14], maximum for $\chi$ close to 19.1° [15], or largest for near-armchair CNTs [16]. Regarding experiments on individual CNTs, *in situ* field emission microscopy by Marchand *et al.* [17] revealed instances of nanotube growths at constant overall rate but made of discrete steps of nanotube rotation. Using *in situ* Raman spectroscopy, Rao *et al.* [18] observed self-exhausting growth kinetics and that the (initial) growth rate was proportional to chiral angle for a set of 9 individual nanotubes. In 2018, Otsuka *et al.* [19] reported an isotope-encoding method with lower time resolution (tens of seconds) but higher throughput (44 nanotubes) which, on the contrary, evidenced a constant growth rate until sudden termination. *In situ* TEM, although very suitable for studying the structure and dynamics of the catalyst particle, is inappropriate for kinetic measurements under real growth conditions due to electron irradiation, low pressure and small field and depth of view.

Here we report high-throughput *in situ* imaging of hundreds of individual CNTs under real growth conditions (at atmospheric pressure, on a substrate) using homodyne polarization microscopy [20,21]. After describing the method, we demonstrate its performances in terms of sensitivity, temporal resolution, field of view and data throughput, and describe the different types of observed behaviors. We then analyze the CNT growth kinetics, starting by the cases displaying one constant growth rate, before moving to the unexpected cases displaying stochastic transitions between different growth rates, and between growth and shrinkage.

Our experimental setup, inspired by the works of Lefebvre *et al.* [22] and Liu *et al.* [20], is schematized in Figure 1. It exploits the optical anisotropy of CNTs to amplify their optical signal relative to that of the substrate, by polarization extinction and by homodyne interference with the field reflected by the substrate [21]. Due to phase effects, the contrast is proportional to the nanotube optical absorption. Experimental details about the setup are given in Supporting Information (SI). We chose iron nanoparticles with ethanol as carbon precursor, one of the most studied systems. ST-cut quartz substrates were used to grow long CNTs aligned along the [100] axis by lattice-oriented growth. We optimized the catalyst concentration to have a sufficiently low density of aligned nanotubes (as assessed by SEM and AFM) to reduce the risk of bundling.

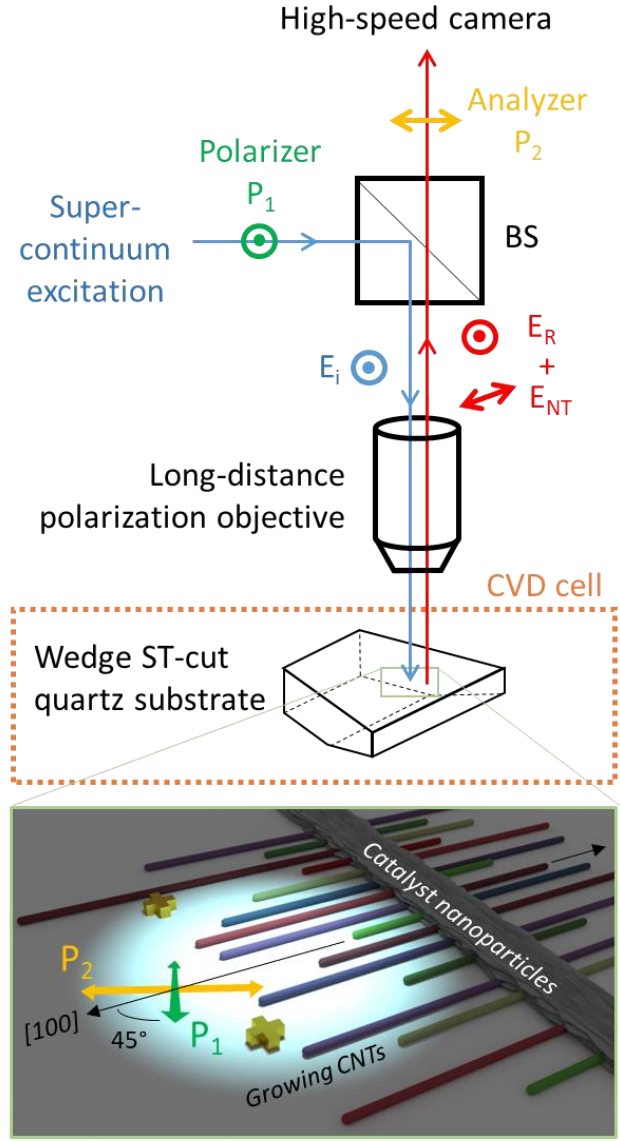

**Figure 1. Principle of the *in situ* optical imaging of individual CNTs during growth by homodyne polarization microscopy.** A supercontinuum white light (400-700 nm) is used to illuminate the inside of a miniaturized CVD cell using a long-distance objective. A ST-cut quartz substrate with a polished wedge (to prevent backside reflection), optical marks and catalyst stripes is used to grow long CNTs (typically 1-100 µm) parallel to its [100] axis. Two polarizers in cross configuration are used to extract the optical signal

of the CNTs growing at an angle of ~ 45° with respect to the polarizers. The signal is further amplified by homodyning with the reflected field and spatially recorded with a high-speed camera.

To improve the video analysis, we have developed a method of rolling-frame correction where each frame is corrected by the frame recorded at a chosen delay δ before, typically between 5 and 20 s. As shown in Fig. 2A and Movie S1, the rolling-frame correction strongly improves the nanotube contrast while highlighting the evolutions during the interval δ. After rolling-frame correction, newly grown CNT sections appear as dark segments whose length divided by δ provides the instantaneous growth rate. It can thus be seen that some nanotubes keep the same growth rate, while others abruptly switch from one constant rate to another constant rate, until sudden termination in both cases. Fig. 2B shows SEM pictures recorded at the same position after the synthesis under SEM conditions allowing either all CNTs (right) or only metallic CNTs (left) to be observed (SEM details are given in SI). As illustrated in Movie S2, we performed a detailed AFM characterization which allowed to observe all nanotubes whatever their electronic type and showed in more than 95 % of cases a one-to-one correspondence (in position and length) between the nanotubes observed in AFM and those observed in the *in situ* video after rolling-frame correction. As detailed in SI (figure S1), this value is in good agreement with the theoretical percentage of CNTs having at least one optical resonance in the experimental range of 400-700 nm. The rolling-frame correction allows resolving the elongation of each individual nanotube as a distinct segment even in the case where CNTs would grow successively at a close position (*i.e.* at less than 1 µm from each other, which is the spatial resolution of our optical setup) as shown in Movie S1.

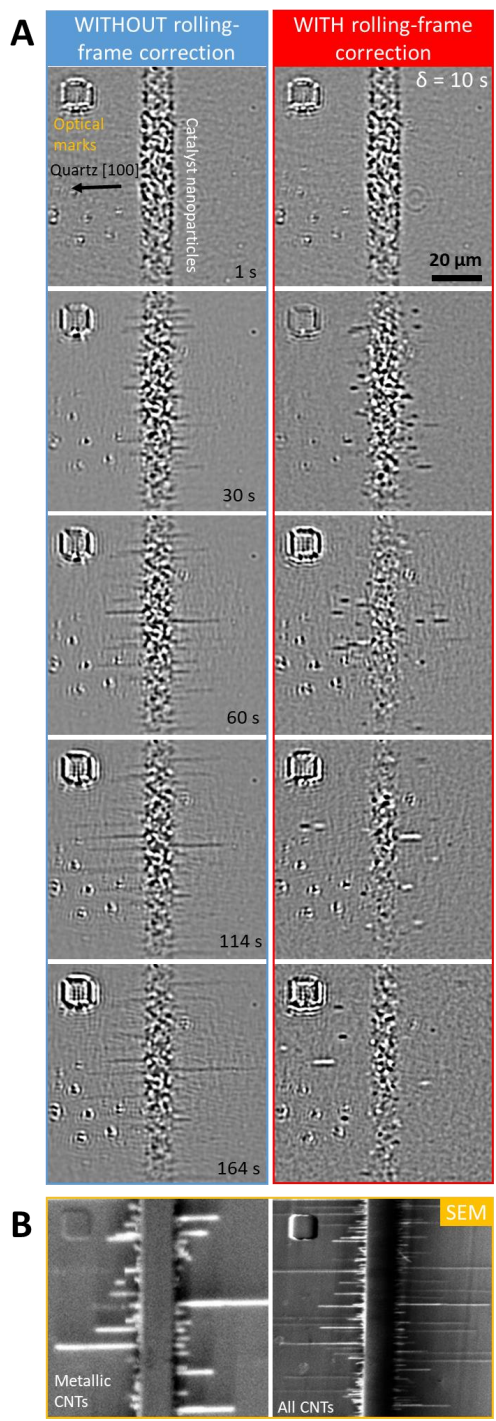

**Fig. 2. *In situ* imaging of CNT growth.** (A) Video snapshots of the same growth sequence without (left) and with (right) rolling-frame correction (delay δ of 10 s): with rolling-frame correction, newly grown CNT sections appear as dark segments, newly etched CNT sections as bright segments and changes of chirality

appear as two consecutive segments with varying contrasts but equal lengths. (B) SEM pictures recorded at the same position after the end of the growth (growth duration: 300 s) under SEM conditions allowing either all CNTs (right) or only metallic CNTs (left) to be observed: comparing SEM pictures and rolling-frame-corrected videos shows that almost all nanotubes in the field of view are individually observed during growth.

Chirality changes can also be clearly observed after rolling-frame correction: as visible in Fig. 2A and movies S1-S5, they are manifested by two consecutive segments with the same length (thus the same rate) moving synchronously. These features indicate a nanotube made of two segments of different chirality that slides on the substrate. The first segment is always dark while the next one can be dark or bright depending on the change in optical absorption. Note that these features demonstrate a base-growth mechanism (*i.e.* the catalyst particle remains fixed while the CNT slides on the monocrystalline substrate) in agreement with previous observations on the same system [19]. Stochastic switches between growth and shrinkage, which are manifested by a switch in both the direction and contrast of a segment, can also be observed.

Of the nearly 700 individual CNTs we have observed, none showed an exponential decrease in growth rate. However, summing the kinetic curves of a large number of individual CNTs does bring a close-to-exponential decay for the average growth rate (Fig. S5). In analogy with radioactive decay, an exponential decay is actually expected when summing the behaviors of individual objects growing with the same constant probability of deactivation. Based on our observations and in agreement with Otsuka *et al.* [19], we conclude that the close-to-exponential decay observed for nanotube ensembles is not a feature of individual CNTs but a consequence of the large distribution of lifetimes.

We first focus on the cases displaying a constant growth rate, as illustrated in Fig. 3A,B. With the Fe/ethanol system, about half of CNTs displayed this behavior. Statistical analysis of 189 such cases evidenced that, under constant growth conditions, both growth rates and lifetimes can vary by up to a factor 30 between CNTs. As shown in Figure 3C, an anti-correlation between growth rate and lifetime is observed: CNTs with a higher growth rate tend to have a shorter lifetime. This indicates that, under these conditions, deactivation is driven like growth by carbon supply, probably through catalyst encapsulation or defect incorporation.

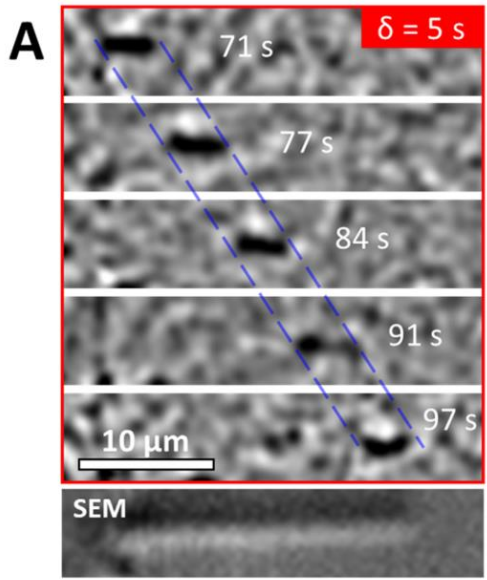

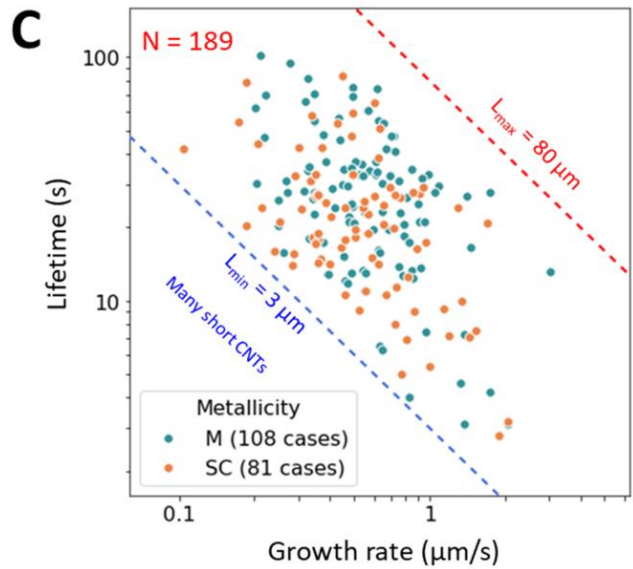

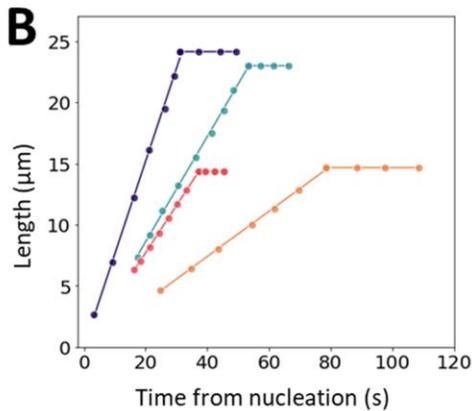

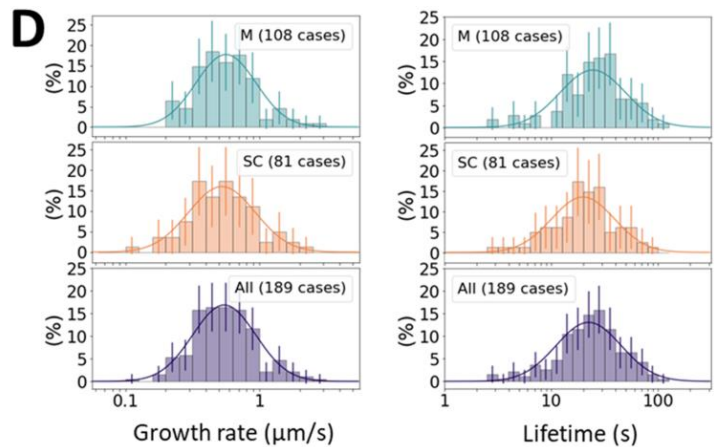

**Figure 3. Nanotube growths at constant rate (representing about half of cases in our standard conditions: T = 850°C, $P_{EtOH}$ = 1600 Pa).** (A) Video snapshots (with rolling-frame correction and delay δ of 5 s) of a CNT growth with constant rate and corresponding SEM picture after growth. (B) Representative kinetic curves (length *versus* time) of CNTs grown at constant rate. (C) Distribution of lifetime *versus* growth rate values. (D) Distributions of growth rates and lifetimes for M-CNTs, SC-CNTs and all CNTs showing no statistical difference in growth rate, lifetime or length between M- and SC-CNTs in the here used conditions.

The M/SC type of these CNTs can be assigned from their SEM contrast (high for M-CNTs, low for SC-CNTs [23]), a method already used by Zhu et al. [8] that we further validated by Raman G-band analyses as detailed in SI. This combined SEM-Raman characterization evidenced 75 % of CNTs fulfilling all the signatures of individual SWCNTs grown on quartz. The other 25 % of cases were assigned as probably not individual SWCNTs but double-walled CNTs (DWCNTs) or two SWCNTs grown at close positions (< 1 µm). In contrast with the report that SC-CNTs grow an order of magnitude faster than M-CNTs [8], our measurements did not reveal any statistical difference in growth rate, lifetime or length between M- and SC-CNTs (Fig. 3D). Faster growth of SC-CNTs should therefore not be considered as a universal behavior but to depend on the conditions: *e.g.* kite-growth, methane and mostly DWCNTs in [8] *versus* lattice-growth, ethanol and mostly SWCNTs in this work.

We now turn to the cases displaying dynamic instability, that is stochastic transitions between different growth rates, and between growth and shrinkage, which represented more than half of CNTs during our observations. We first checked that these changes were not caused by uncontrolled fluctuations of the CVD conditions: first, the temperature was well stabilized before introducing ethanol and remained stable at +/-1 °C during experiments; second, the Ar/ethanol mixture was equilibrated in a parallel line before being switched into the CVD cell; third, despite an initial increase due to ethanol decomposition, the $H_2O$ concentration remained stable until the end of the experiment. Most importantly, based on the analysis of a hundred *in situ* videos, the transitions occur without any correlation, spatial or temporal, between CNTs (see Movies S3-5).

Most transitions in growth rate (about 75 %, *i.e.* 201 out of 266) occur without a change in contrast in both *in situ* optical imaging and *ex situ* SEM (Fig. 4A,B). We performed a multi-wavelength Raman study which confirmed that the resonant Raman features (laser resonance energy, RBM position, G-band shape) along these CNTs remain unchanged for the vast majority of them (86 %, *i.e.* 37 out of 43 Raman-resonant CNTs) despite the rate change (Figure S6): this indicates that, in the vast majority of such cases, the rate changes occur without a change in chirality. In contrast, about 25% of rate change events (65 out of 266) were correlated with a change in optical contrast, that is to a change in chirality (Fig. 4C,D). Such chirality changes were always corroborated by a change in resonant Raman features (Figure S7), and, in the case of a M/SC transition, by a change in SEM contrast (Fig. 4C), at the expected position along the nanotube.

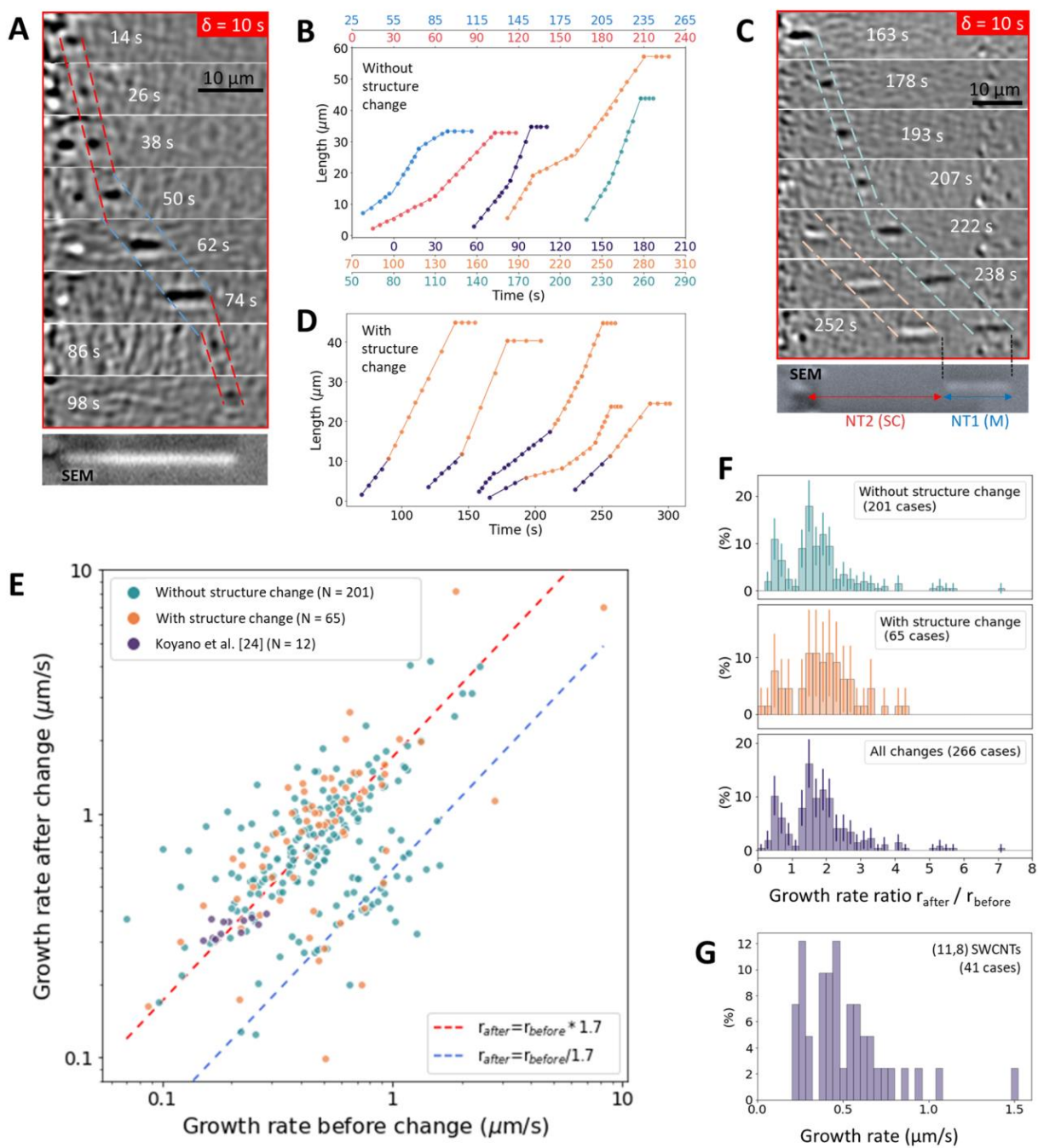

**Fig. 4. Stochastic switches of growth rates (representing more than half of cases in our standard conditions: T = 850°C, P_EtOH = 1600 Pa).** (A, B) Changes of growth rate with no change of CNT structure: (A) video snapshots and corresponding SEM picture; (B) examples of kinetic curves (for easier viewing, each curve has been given its own time axis with the same color code). (C, D) Changes of growth rate

correlated with a change of CNT structure as evidenced by the changes in optical absorption, SEM contrast and Raman resonance: (C) video snapshots and corresponding SEM picture taken after growth; (D) kinetic curves displaying a change in growth rate correlated with a change from a first chirality (blue points) to a second one (orange points). Both videos were corrected with the rolling-frame correction ($\delta = 10$ s). (E) Correlation plot of the growth rates before and after change (N is the number of cases) displaying two elongated clouds with an average proportionality factor of 1.7 for both rate increase and decrease events. (F) Distributions of the growth rate ratio $r_{after}/r_{before}$ with and without structure change, and for all cases. (G) Distribution of growth rates for (11,8) SWCNTs displaying a multimodal shape despite the constant chirality and growth conditions.

The correlation plot of the growth rates before and after change for 266 transitions is shown in Figure 4E. The graph reveals two distinct elongated clouds, one for rate increases and one for rate decreases. Both clouds evidence the same proportionality factor of about 1.7 (or $1/1.7 = 0.6$) between consecutive rates: this indicates that the two clouds correspond to transitions between the same rate levels but in opposite directions. Importantly, the same correlation is observed with or without chirality change (orange and green points, respectively). In both cases, most transitions (201 out of 266, *i.e.* 76 %) correspond to a growth rate increase, which also supports that the same mechanism is at play with and without chirality change. Figure 4F precisely shows the distribution of the ratio of growth rates after and before change ($r_{after}/r_{before}$): two distinct massifs can be seen at $r_{after}/r_{before} = 1.7$ and 0.6 (*i.e.* 1/1.7) for growth rate increases and decreases, respectively. Note that the statistical resolution does not allow excluding a possible substructure in each massif (Fig. S8).

To study the influence of chirality, we performed a Raman characterization on 230 individual CNTs (corresponding to a total of 350 CNT segments) displaying a rate change or not. This allowed us to assign 41 CNTs or CNT segments to (11,8) SWCNTs thanks to their appropriate position in the Kataura plot and high occurrence in our samples (Fig. S4). As shown in Figure 4G, the growth rate distribution of (11,8) SWCNTs is clearly not monomodal, in contradiction with current theories: it displays two main peaks at about 0.25 µm/s and 0.5 µm /s, with a possible substructure in the second peak. This provides a direct evidence that, even at constant chirality and under constant conditions, CNTs do not display a single growth rate but a few discrete rates. Strikingly, Koyano *et al.* reported a similar factor of 1.7 for the growth rates of individual CNTs before and after an interruption with $Ar/H_2/H_2O$ [24]: as shown in Fig. 4E, the data of Koyano *et al.* fit well within our larger set of data. This suggests a similar mechanism despite different experimental conditions.

We finally move to CNTs displaying stochastic switches between growth and shrinkage as illustrated in Fig. 5A. Shrinkage has a noticeable impact on the yield since approximately 18% of CNTs exhibited at least one switch between growth and shrinkage under the studied conditions. From the cases displaying chirality change followed by shrinkage, it can be deduced that the shrinkage proceeds by carbon etching at the interface with the catalyst particle (Movie S6). During our experiments, etching appears to be caused by ethanol or its sub-products (*e.g.* $H_2O$) since we observed that shrinkage stopped when interrupting the ethanol supply. Background traces of $H_2O$ and $O_2$, which are commonly present during CNT growth by CVD (~10 ppm $O_2$ and ~50 ppm $H_2O$ in our experiments), may also play a role. Despite the growth rate fluctuations, the data suggest a positive correlation between growth and etching rates (figure S9). As visible in Fig. 5B,C, there is often a pause between growth and shrinkage. Pauses from a few seconds to 150 s can also be observed between two sequences of growth or shrinkage. We cannot exclude that these pauses correspond to extremely slow rates outside the reach of our spatial resolution (~1 µm). A few extreme cases even displayed several reversible switches between growth, pause, and shrinkage (Figure 5C), thus confirming the stochastic nature of the switches.

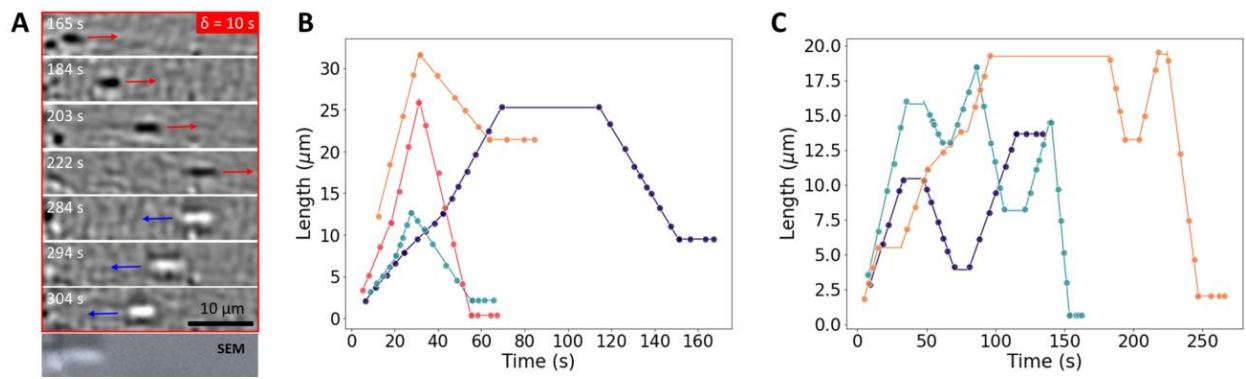

**Fig. 5. Stochastic switches between growth and shrinkage: about 18 % of CNTs displayed at least one such switch in our standard conditions.** (A) Video snapshots (with rolling-frame correction with a delay δ of 10 s) illustrating a switch from growth to shrinkage (which are manifested by a switch in both the direction and contrast of a CNT segment), and corresponding SEM picture after growth. (B,C) Examples of kinetic curves with one switch (B) or several switches (C) between growth and shrinkage.

We now discuss the possible mechanisms behind the dynamic instabilities revealed by our observations. We note that these instabilities bear analogies with similar phenomena displayed by other nanoscale systems, such as the dynamic instability of microtubule growth [25,26] or the catalytic growth of nanowires for which oscillations of the growth interface [27], catalyst jumping between different facets [28] and fluctuations in growth rates [29] have been reported. We have already ruled out local fluctuations in pressure or feedstock composition as a possible cause of these instabilities since the behaviors of neighboring nanotubes are uncorrelated. Catalyst ripening affects the catalyst particle sizes during CNT growth [30] but cannot induce reversible switches between discrete states due to its continuous and irreversible nature. Having eliminated other hypotheses, the explanation must be related to structural or configurational switches of the nanotube edge or catalyst nanoparticle. The growth rate of 3D and 2D crystals depends on the crystal facets or edges, but the structure of the nanotube edge during growth and how it depends on chirality, catalyst and growth conditions are unknown. In line with the role of edge entropy to stabilize chiral CNTs [31], an attractive hypothesis for growth rate switches is that CNT edges switch between different stable configurations (*e.g.* with mostly armchair sites or mostly zigzag sites). In such a hypothesis, the switching frequency and the proportionality factor between slow and fast rates should depend on nanotube chirality. Fluctuations of the crystalline orientation of the catalyst nanoparticle, which are commonly observed by *in situ* TEM [32–37], may also play a role.

However, these hypotheses cannot explain switches between growth and shrinkage which require a change in carbon chemical potential at the nanotube-catalyst interface. In this case, assuming that catalyst nanoparticle switches between two phases with different catalytic activities for precursor decomposition (*e.g.*, between metal and carbide) could provide an explanation. *In situ* XPS [38] and TEM [39] during CNT growth have actually shown that catalyst nanoparticles can adopt different crystal phases with different catalytic activity (*e.g.*, Fe-α, Fe-γ, $Fe_3C$, $Fe_5C_2$ in the case of iron). Sharma *et al.* notably reported deactivation events correlated with phase changes from $Fe_3C$ to $Fe_5C_2$ [39] or from $Co_2C$ to $Co_3C$ [34]. Since the stability of these phases depends on carbon concentration, phase transitions may be caused by an imbalance between carbon supply and consumption.

To assess these hypotheses, we tested nickel, which does not form stable carbides, instead of iron. We did observe rate change events with Ni (Movie S7), although with a lower frequency than with Fe, but we did not observe shrinkage events. This supports that switches between growth and shrinkage are specific to catalysts displaying different possible phases during growth, like iron. In contrast, rate changes appear as a more general behavior potentially related to configurational switches of the nanotube edge. Further experimental and theoretical investigations are needed to elucidate the origin of these instabilities.

Our observations of a large number of individual CNTs under real growth conditions reveal processes more complex than described by current models. In particular, they contradict the standard assumption of a unique growth rate for a given chirality. New growth models should be built to account for the growth rate instabilities revealed by our observations. The fact that the highest selectivity values are obtained with solid-state catalysts might indicate that their actual role is to limit these fluctuations and stabilize the growth rate. The understanding and control of these instabilities therefore seem paramount for the rational design of highly selective CNT growth methods. Interestingly, instabilities between growth, chirality change and shrinkage may offer opportunities of chirality selection based on a dynamic exploration of the chirality space over multiple cycles. Finally, we emphasize that *in situ* homodyne polarization microscopy is not limited to CNTs but can be applied to many nanostructures (0D, 1D or 2D) in different environments (vacuum, gas or liquid) as long as the nanostructures (or their edges) scatter light with sufficient efficiency and polarization change.

**List of Supporting information:**

Materials and Methods: catalyst preparation, CNT growth, *in situ* optical microscopy, video processing and analysis, classification of CNT growth kinetics, CNT characterization.

Figures S1-S10

Movie S1: comparison between videos of the same growth sequence with and without rolling-frame correction

Movie S2: comparison between *in situ* video and post-growth AFM and SEM images

Movies S3-5: additional examples of *in situ* videos with rolling-frame correction

Movie S6: examples of cases displaying a chirality change followed by shrinkage

Movie S7: examples of sudden changes of growth rates with Ni as catalyst

**Acknowledgments:** The authors acknowledge the support of the *Agence Nationale de la Recherche* (grant ANR-20-CE09-0007-01). This project has received financial support from the CNRS through the MITI interdisciplinary programs. This material is based upon work supported by the Air Force Office of Scientific Research under award number FA9550-17-1-0027. VJ acknowledges the support of the *Institut Universitaire de France*.

**Author contributions:** V.J. designed and coordinated the research. T.M. and L.M. conceived and built the optical setup. H.N.T. developed the methods for *in situ* optical imaging during CNT growth. S.T. prepared the catalyst samples and performed SEM characterization. V.P. performed the *in situ* optical imaging and the Raman characterization, developed the protocols of data treatment and analyzed the experimental data. M.O. and R.P. performed the AFM characterization. C.B. contributed to the data analysis and the manuscript writing. V.P. and V.J. co-wrote the manuscript. All authors contributed to critical discussions of the manuscript.